\documentclass[aps,prb,superscriptaddress,twocolumn,nofootinbib]{revtex4}
\usepackage[english]{babel}
\usepackage{amssymb}
\usepackage{amsmath}
\usepackage{graphicx}
\begin{document}
%\fontsize{11pt}{12pt}\selectfont % for font setting, change the first number

\title{Robust recipe for low-resistance ohmic contacts to a two-dimensional
electron gas in a GaAs/AlGaAs heterostructure}

\author{M.~J.~Iqbal\footnote{javaid2k@gmail.com}}
\affiliation{Zernike Institute for Advanced Materials, Nijenborgh 4, University of Groningen, NL-9747AG Groningen, The Netherlands}
\affiliation{Centre of Excellence in Solid State Physics, University of the Punjab, QAC, Lahore-54570, Pakistan}
\author{D.~Reuter}\thanks{Now at Department of Physics, University Paderborn, Warburger Stra{\ss}e 100, 30098 Paderborn, Germany.}
\author{A.~D.~Wieck}
\affiliation{Angewandte Festk\"{o}rperphysik, Ruhr-Universit\"{a}t
Bochum, D-44780 Bochum, Germany}
\author{C.~H.~van~der~Wal}
\affiliation{Zernike Institute for Advanced Materials, Nijenborgh 4, University of
Groningen, NL-9747AG Groningen, The Netherlands}

\date{version \today}
%%%%%

\begin{abstract}  %% uncomment for complete compile
%\fontsize{12pt}{12pt}\selectfont
The study of electron transport in low-dimensional systems is of importance, not only from a fundamental point of view, but also for future electronic and spintronic devices.
In this context heterostructures containing a two-dimensional electron gas (2DEG) are a key technology. In particular GaAs/AlGaAs heterostructures, with a 2DEG at typically 100~nm below the surface, are widely studied.
In order to explore electron transport in such systems, low-resistance ohmic contacts are required that connect the 2DEG to macroscopic measurement leads at the surface. Here we report on designing and measuring a dedicated
device for unraveling the various resistance contributions in such contacts, which include pristine 2DEG series resistance, the 2DEG resistance under a contact, the contact resistance itself, and the influence of pressing a bonding wire onto a contact. We also report here a robust recipe for contacts with very low resistance, with values that do not change significantly for annealing times between 20 and 350 sec, hence providing the flexibility to use this method for materials with different 2DEG depths. The type of heating used for annealing is found to strongly influence the annealing process and hence the quality of the resulting contacts.

\textbf{Keywords}: III-V semiconductors, Ohmic contacts, heterostructures, 2DEG, Electrical properties
%\rule{\textwidth}{0.7pt}% for putting horizontal line
\begin{center}
\line(1,0){450}
\end{center}

\end{abstract} %% uncomment for complete compile

%\pacs{xxx.xxx.xxx}

%\clearpage %uncomment originally

\maketitle %comment for full compile
\clearpage %comment for full compile

%%%%%%%%%%%%%%%%%%%%%%%%%%%%%%%%%%%%%%%%%%%%%%%%%%%%%%%%%%%%%%%%%%%%%%%%%
%%% INTRODUCTION %%%%%%%%%%%%%%%%%%%%%%%%%%%%%%%%%%%%%%%%%%%%%%%%%%%%%%%%
%%%%%%%%%%%%%%%%%%%%%%%%%%%%%%%%%%%%%%%%%%%%%%%%%%%%%%%%%%%%%%%%%%%%%%%%%

\renewcommand\thesection{\arabic{section}}% For numering to the sections

\section{\label{8sec:intro}Introduction}
The two-dimensional electron gas (2DEG) is of interest for the study of low-dimensional systems,  and high-mobility 2DEGs can be realized in epitaxially grown GaAs/${\rm Al}_{x}{\rm Ga}_{1-x}{\rm As}$ heterostructures \cite{9Beenakker1991,9Kouwenhoven1998}. For performing electrical transport experiments on these systems ohmic contacts to the 2DEG are very important, and these can be realized by annealing samples after a metal alloy has been deposited on the surface at the intended contact areas. Commonly, an alloy consisting of AuGe/Ni/Au is used \cite{9Baca1997}. The annealing times and temperatures that give the lowest contact-resistance values are different for different 2DEG depths. In our previous work \cite{IqbalSST2013,PhDThesis} we optimized such a recipe for annealing in a glass-tube oven. For the study presented here, we used different annealing conditions, namely annealing in a Rapid Thermal Annealer (RTA). Surprisingly, annealing with the RTA gives much lower contact-resistance values with similar values for a very wide range of annealing times. These results are attributed to the exact heating profile as a function of time during annealing, and the process that we followed for cleaning the samples. We already applied our recipes for ohmic contacts in our studies of quantum point contacts in a 2DEG \cite{IqbalNat2013,IqbalJAP2013}.

The AuGe/Ni/Au material was first used by Braslau
\textit{et al.} \cite{9Braslau1967} to make an ohmic contact to \mbox{n-GaAs} in 1967. Subsequent studies aimed at improving such contacts and understanding the annealing mechanisms \cite{9Ogawa1980,9Braslau1981,9Lee1981,9Heiblum1982,9Kuan1983,9Braslau1983,9Braslau1986,9Procop1987,9Waldrop1987,9Bruce1987,9Shappirio1987,9Relling1988,9Weizer1988,9Lumpkin1996}. Later on, with the increasing importance of the 2DEG in a GaAs/${\rm Al}_{x}{\rm Ga}_{1-x}{\rm As}$ heterostructure, research focussed on making ohmic contacts to the buried 2DEG \cite{9Zwicknagl1986,9Higman1986,9Rai1988,9Jin1991,9Taylor1994,9Messica1995,9Taylor1998,9Raiser2005,9Saravanan2008}. Despite these extensive studies, a model was missing that could predict the optimal annealing times and temperatures for different depths of the 2DEG. In our previous work we developed such a model \cite{IqbalSST2013,PhDThesis}. In order to understand the annealing mechanism further, we studied the contact-resistance values as a function of circumference and area of the ohmic contacts. However, no clear dependence on circumference or area was found \cite{IqbalSST2013,PhDThesis}, in part because of lack of information on whether the 2DEG square resistance under an ohmic contact changes during the annealing with respect to the square resistance of pristine 2DEG. In addition, it was unknown whether pressing a bonding wire on a contact influences the resistance, as it can possibly rapture the
ohmic contact layer over a significant area. Here we report on studying these
questions. We designed a dedicated device structure that allowed us
to study the contact resistance with various resistance contributions in more detail, and with
different measurement methods (3-point measurement, 4-point measurement and the Transmission Line Method (TLM) \cite{8Berger1972TLM,8Reeves1982TLM}).

All the results presented in this paper are from samples annealed
in a rapid thermal annealer (RTA), unlike our previous work
where the hot gas flow in a glass-tube furnace was used for
annealing the samples. In the present study we focused on changes in
the mentioned contact resistance contributions as a function of
annealing time (at fixed annealing temperature and depth of the
2DEG). However, the contact resistance values that we found were, for
all samples, so low that we could not obtain sufficient measurement
accuracy for determining a dependence on annealing time for the
square resistance of 2DEG under an ohmic contact. Instead, the value of
these results is that we found a very robust recipe for low
contact-resistance values, with values that do not change significantly for
annealing times between 20 and 350 sec. This kind of robust recipe
is useful for annealing contacts on 2DEG materials with different
2DEG depths where one would need a large margin for suitable
annealing times. Our results also show that the details of how
the sample is heated have a strong influence on the annealing
mechanism.

\section{\label{8sec:Experimental}Experimental details}

For the present study we used two different wafers, one with the 2DEG at 60~nm depth (wafer I, purchased from Sumitomo Electric Industries, Inc.) and one with the 2DEG at 180~nm depth (wafer II, grown by our team in Bochum). Unless mentioned otherwise, we present results of devices that were fabricated with wafer I. The study of devices from wafer II were less extensive but we will mention the results that are relevant. All the measurements were performed in a liquid helium vessel at 4.2~K.

Wafer I was a GaAs/${\rm Al}_{0.27}{\rm Ga}_{0.73}{\rm As}$ heterostructure. The layer sequence of the
heterostructure was as follows (top to bottom): a 5~nm $n$-GaAs cap, 40~nm ${\rm Al}_{0.27}{\rm Ga}_{0.73}{\rm As}$ \mbox{$n$-doped} with Si at $2.0 \times 10^{18}~ \text{cm}^{-3}$, a
15~nm nominally intrinsic ${\rm Al}_{0.27}{\rm Ga}_{0.73}{\rm As}$ spacer
layer, and a 800~nm GaAs layer. The 2DEG is located at the interface
of the AlGaAs spacer layer and the next GaAs layer. The 2DEG density
and mobility at $4.2$ K were $n_{2D} = \text{3.30} \times 10^{15}~
\text{m}^{-2}$ and $\mu_{2D} = 19.8~\text{m}^2/\text{Vs}$,
respectively. Wafer II was a similar GaAs/${\rm Al}_{0.35}{\rm Ga}_{0.65}{\rm As}$ heterostructure with the layer sequence (top to bottom): a 5~nm $n$-GaAs cap, 70~nm ${\rm Al}_{0.35}{\rm Ga}_{0.65}{\rm As}$, 70~nm ${\rm Al}_{0.35}{\rm Ga}_{0.65}{\rm As}$ \mbox{$n$-doped} with Si at $\sim 1.0 \times 10^{18}~ \text{cm}^{-3}$, 35~nm ${\rm Al}_{0.35}{\rm Ga}_{0.65}{\rm As}$, and 650~nm GaAs. It had  $n_{2D} = \text{1.93} \times 10^{15}~\text{m}^{-2}$ and $\mu_{2D} = 33.3~\text{m}^2/\text{Vs}$.

Several cleaning steps during ohmic contact fabrication are very important for getting low-resistance ohmic contacts. The cleaning process is done before starting the ohmic contact fabrication. The samples are first cleaned in acetone, and then in iso-propyl-alcohol, while keeping the sample in an ultrasonic bath on a low power. The samples are then visually inspected and only samples that appear fully clean are used. We observed that contaminated samples show high resistance values and results that cannot be reproduced.

The size of the ohmic contacts was 200 by 200~${\rm \mu m^{2}}$ and
they were patterned with electron-beam lithography. For the ohmic
contacts, layers of  AuGe in eutectic-composition (12 wt$\%$ Ge,
150~nm), Ni (30~nm) and Au (20~nm) were deposited subsequently by
electron-beam evaporation. The contacts were annealed at
450~$^{\circ}$C in the rapid thermal annealer (RTA, model Jipelec
Jet 150) for various times. Annealing took place in a clean ${\rm
N_2}$ flow (600~sccm) to avoid oxidation and material vapors
adhering back onto the sample. During annealing the functional
sample surface was directly facing the RTA heating lamps.

%%%%%%%%%%%%%%%%%%%%%%%%%%%%%%%%%%%%%%%%%%%%%%%%%%%%%%%%%%%%%%%%%%%%%%%%%%%
%RTA profile
%ChapterOhmLowRes/
\begin{figure}[h!]%%h!
\centering
\includegraphics[width=\columnwidth]{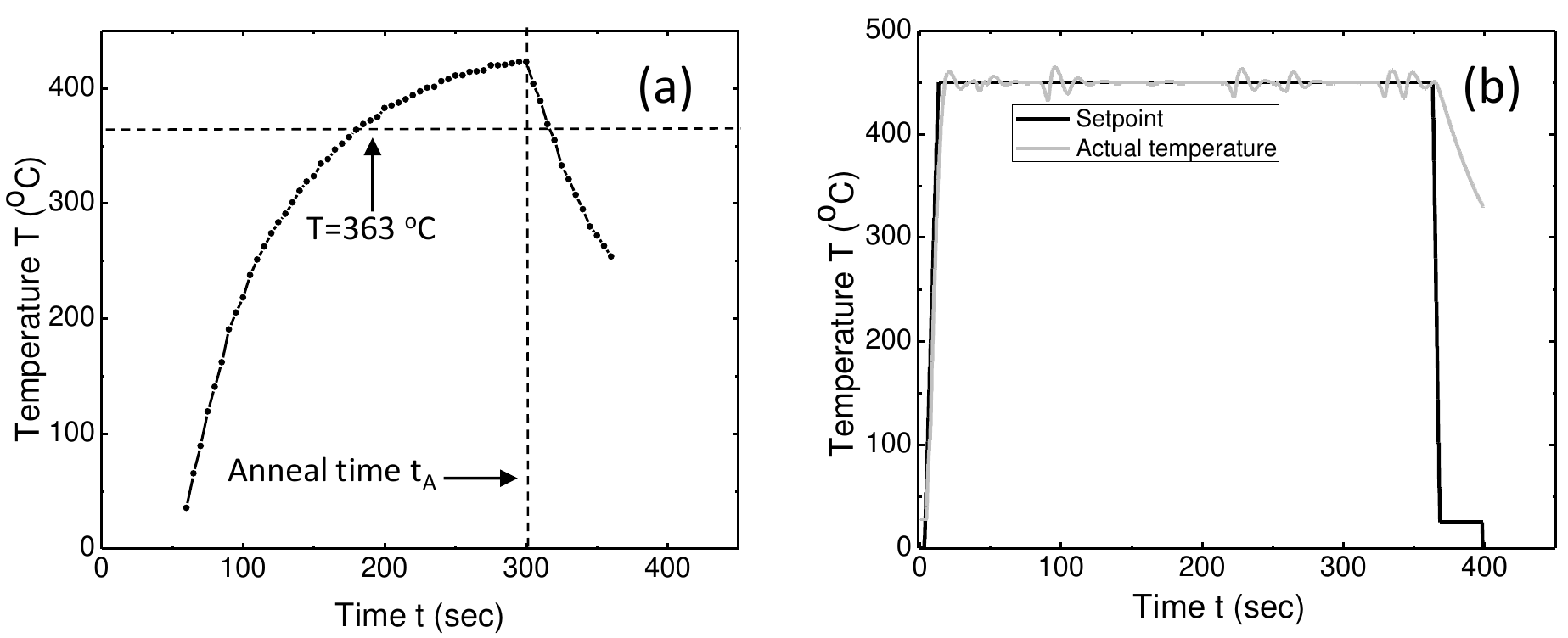}
\caption{Temperature of a thermocouple in close contact with the
sample as a function of time during the annealing process in the
glass-tube furnace \textbf{(a)} and rapid thermal annealer (RTA)
\textbf{(b)}.} \label{8Fig:RTAprofile}
\end{figure}
%%%%%%%%%%%%%%%%%%%%%%%%%%%%%%%%%%%%%%%%%%%%%%%%%%%%%%%%%%%%%%%%%%%

The annealing temperature profiles for the glass-tube oven (used in our previous work \cite{IqbalSST2013}) and the RTA are shown in Fig.~\ref{8Fig:RTAprofile}a,b. For the glass-tube oven, the sample is brought into a pre-heated oven and the temperature rise of the
sample holder to the AuGe-eutectic temperature (363$^{\circ}$C)
takes a few minutes (not easily controllable). For the RTA, on the
other hand, the temperature ramp rate can be controlled and the time
for reaching the set temperature was set at the much shorter (but
for RTA heating typical) value of 5~sec. This is started with the
sample already in the oven. The black and light gray lines in
Fig.~\ref{8Fig:RTAprofile}b are for the set temperature and the
actual temperature as measured by the thermocouple attached to the
surface where sample is placed for annealing. All the results
presented in this paper are for samples annealed in the RTA.

%%%%%%%%%%%%%%%%%%%%%%%%%%%%%%%%%%%%%%%%%%%%%%%%%%%%%%%%%%%%%%%%%%%%%%%%%%%
%RTA Scheme
\begin{figure}[h!]%%h!
\centering
\includegraphics[width=0.85\columnwidth]{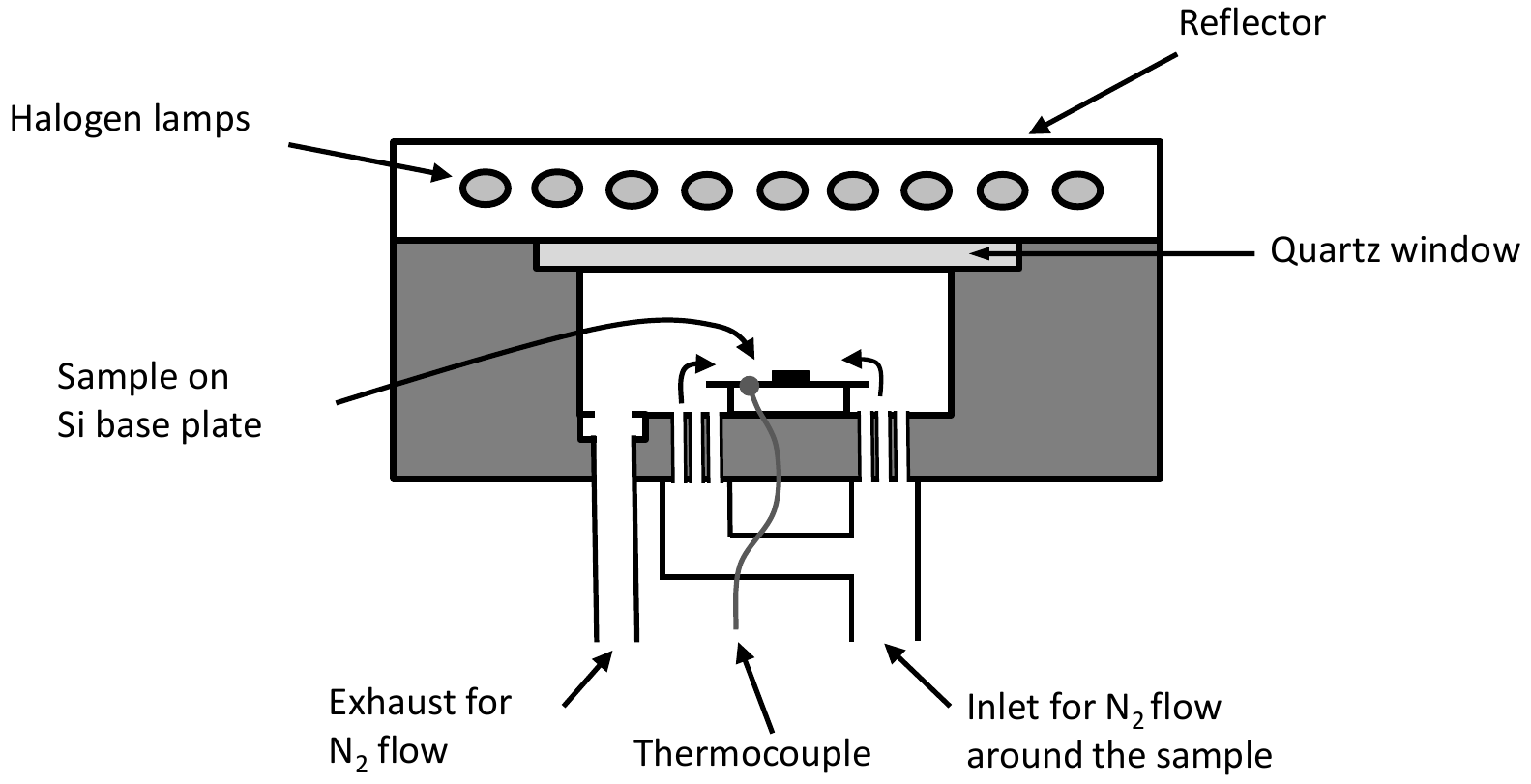}
\caption{Simplified scheme of the RTA annealing chamber.} \label{8Fig:RTAScheme}
\end{figure}
%%%%%%%%%%%%%%%%%%%%%%%%%%%%%%%%%%%%%%%%%%%%%%%%%%%%%%%%%%%%%%%%%%%

Fig.~\ref{8Fig:RTAScheme} shows a schematic of the RTA annealing
chamber. The heating sources are the halogen lamps that transmit
radiation through a quartz window above the sample surface. The
sample is placed on a Si base plate and a thermocouple is attached
to this plate for measuring the annealing temperature during the
process. The flow of ${\rm N_{2}}$ gas is maintained during the
entire annealing process and cool down.

\section{\label{8sec:design}Device design, measurement schemes and methods}

%%%%%%%%%%%%%%%%%%%%%%%%%%%%%%%%%%%%%%%%%%%%%%%%%%%%%%%%%%%%%%%%%%%%%%%%%%
%Optical image of device.

\begin{figure}[h!]%%h!
\centering
\includegraphics[width=\columnwidth]{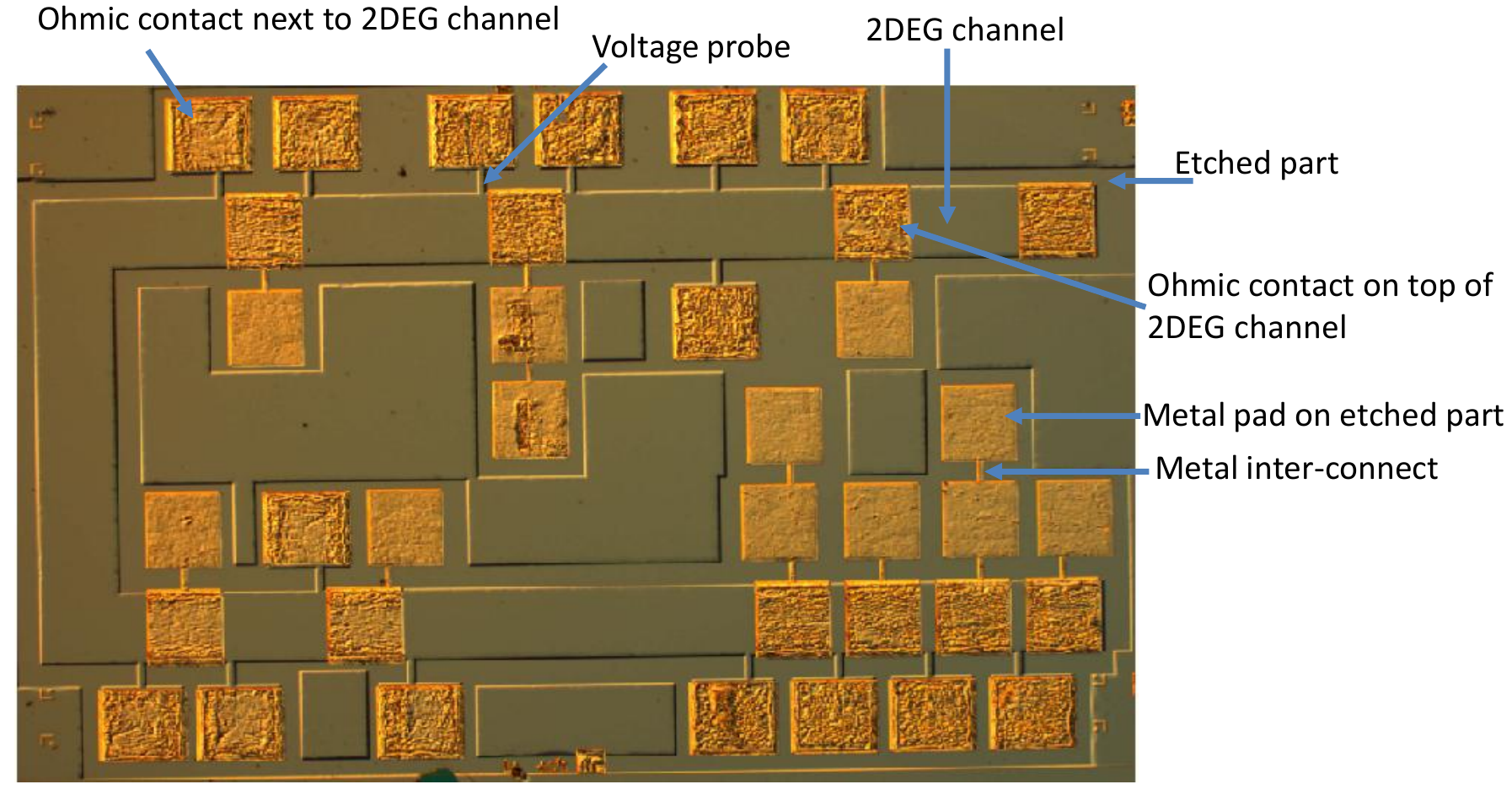}
\caption{Optical microscope image of a device showing etched mesa
regions and deposited contacts. The shallow part of the mesa is
wet-etched at places where the design requires a boundary to a 2DEG
region. The ohmic contacts and metal pads (all 200 by 200~${\rm \mu
m^{2}}$) are deposited and annealed on 2DEG and etched surface
respectively so that the latter are separated from the 2DEG.}
\label{8Fig:Device}
\end{figure}
%%%%%%%%%%%%%%%%%%%%%%%%%%%%%%%%%%%%%%%%%%%%%%%%%%%%%%%%%%%%%%%%%%%
%%%%%%%%%%%%%%%%%%%%%%%%%%%%%%%%%%%%%%%%%%%%%%%%%%%%%%%%%%%%%%%%%%%
%%Scheme of device.
\begin{figure}[h!]%%th!
\centering
\includegraphics[width=0.8\columnwidth]{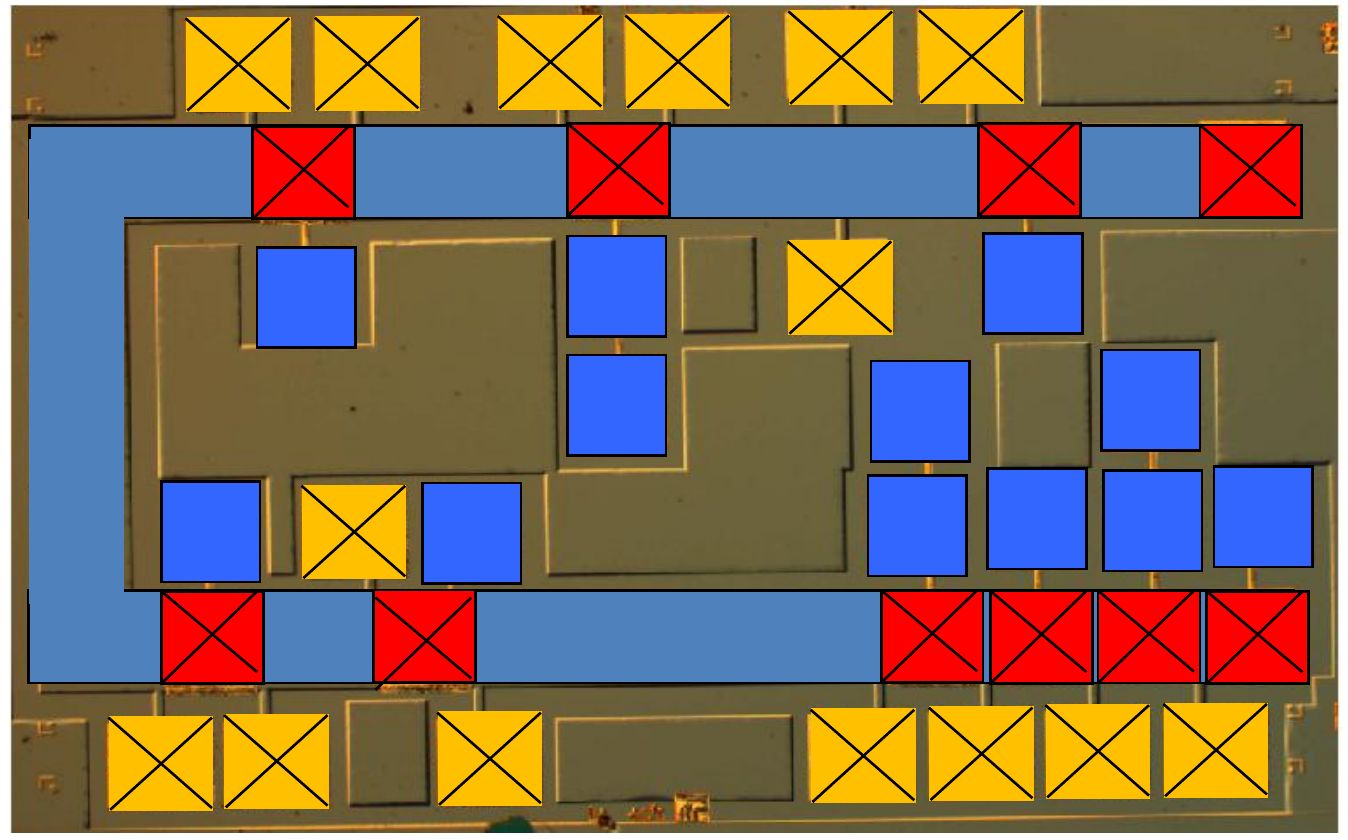}
\caption{Color scheme to highlight different parts of the device.
The U-shaped blue part is a 200~${\rm \mu m}$ wide 2DEG channel. The
red squares are ohmic contacts on top of the main 2DEG channel and
the yellow squares are ohmic contacts on the side of the channel
that only serve as voltage probes. The dark blue squares are metal
contact pads on etched wafer areas (all contacts are 200 by
200~${\rm \mu m^{2}}$).} \label{8Fig:DeviceScheme}
\end{figure}
%%%%%%%%%%%%%%%%%%%%%%%%%%%%%%%%%%%%%%%%%%%%%%%%%%%%%%%%%%%%%%%%%%%

\begin{figure}[!] %%th!
\centering
\includegraphics[width=0.8\columnwidth]{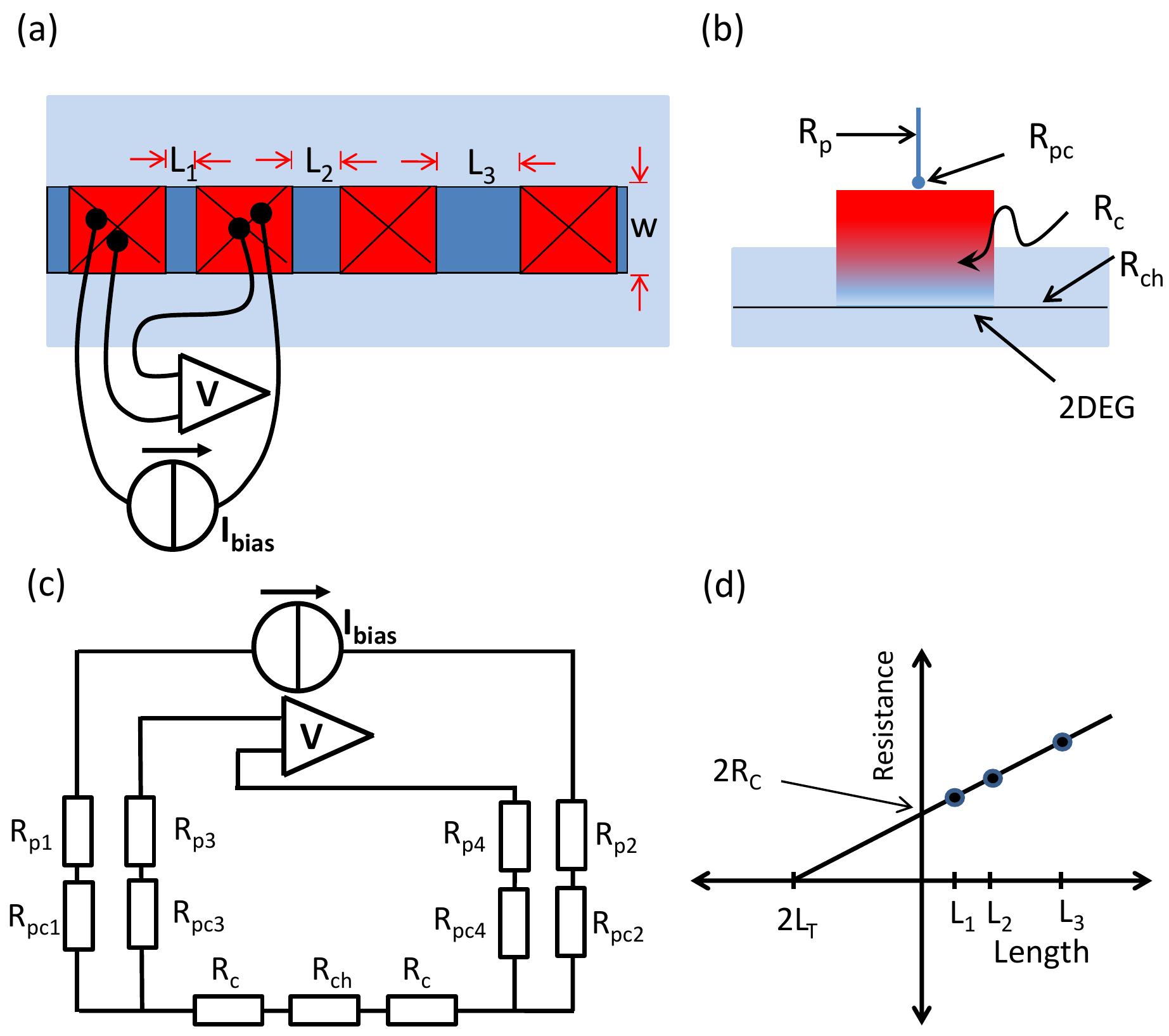}
\caption{Transmission Line Method (TLM) measurement scheme.
\textbf{(a)} Ohmic contacts are made over the full width of a 2DEG
strip with increasing distance $L_i$ between adjacent contacts. A
4-point measurement is used for determining the resistance
$R_{total}$ for each segment. \textbf{(b)} A side view on a contact
showing the various resistance contributions (see main text for
details). \textbf{(c)} A circuit diagram for the $4$-point scheme
with the resistance contributions from panel (b). \textbf{(d)} A
schematic plot of TLM measurement results (see main text for
details).} \label{8Fig:TLM}
\end{figure}

We designed a dedicated device structure for being able to study the
different contact-resistance contributions with different
measurement methods. An optical image of a fabricated device is
shown in Fig.~\ref{8Fig:Device}. A 200~${\rm \mu m}$ wide 2DEG
channel is defined (U-shaped) by wet etching such that a homogeneous
current flow can be applied through the defined strip. Different
parts of the devices are labeled. The ohmic contacts are realized on
top of the channel as well as on the sides of the channel. The
available distances $L_i$ (defined as in Fig.~\ref{8Fig:TLM}a)
between the contacts on top of the channel are 40, 260, 460, 660,
and 860~${\rm \mu m}$. The effect of annealing times on the pristine
2DEG square resistance, and on the full contact resistance and 2DEG
square resistance under ohmic contacts are determined using the ohmic
contacts deposited on top of the 2DEG channel. The ohmic contacts
deposited on the sides are connected to the 2DEG channel via narrow
2DEG strips (20~${\rm \mu m}$ wide) and serve as voltage probes
(used when measuring the resistance of the contacts, the resistance
of pristine 2DEG and the resistance of 2DEG under contacts). Metal
pads and metal inter-connects between pads are deposited on etched
parts of the device. These metal pads are connected to the ohmic
contacts via metal inter-connects and used for measuring the ohmic
contact resistances without directly bonding on the ohmic contacts
themselves (the resistance contributions from metal pads and
inter-connects are subtracted in this case). By comparing values
measured with this bonding scheme to a subsequent measurement with
bonding directly on top of the ohmic contacts, the influence of
pressing a bonding wire on an ohmic contact can be determined. The
color scheme in Fig.~\ref{8Fig:DeviceScheme} further illustrates the
various device parts more clearly. The long 2DEG channel is shown
with a light blue color. The ohmic contacts are shown as crossed
color squares, with contacts on top of the 2DEG channel in red and
side contacts in yellow. The metal pads on etched regions are shown
as dark blue squares.

\begin{figure}[h!]
\centering
\includegraphics[width=0.7\columnwidth]{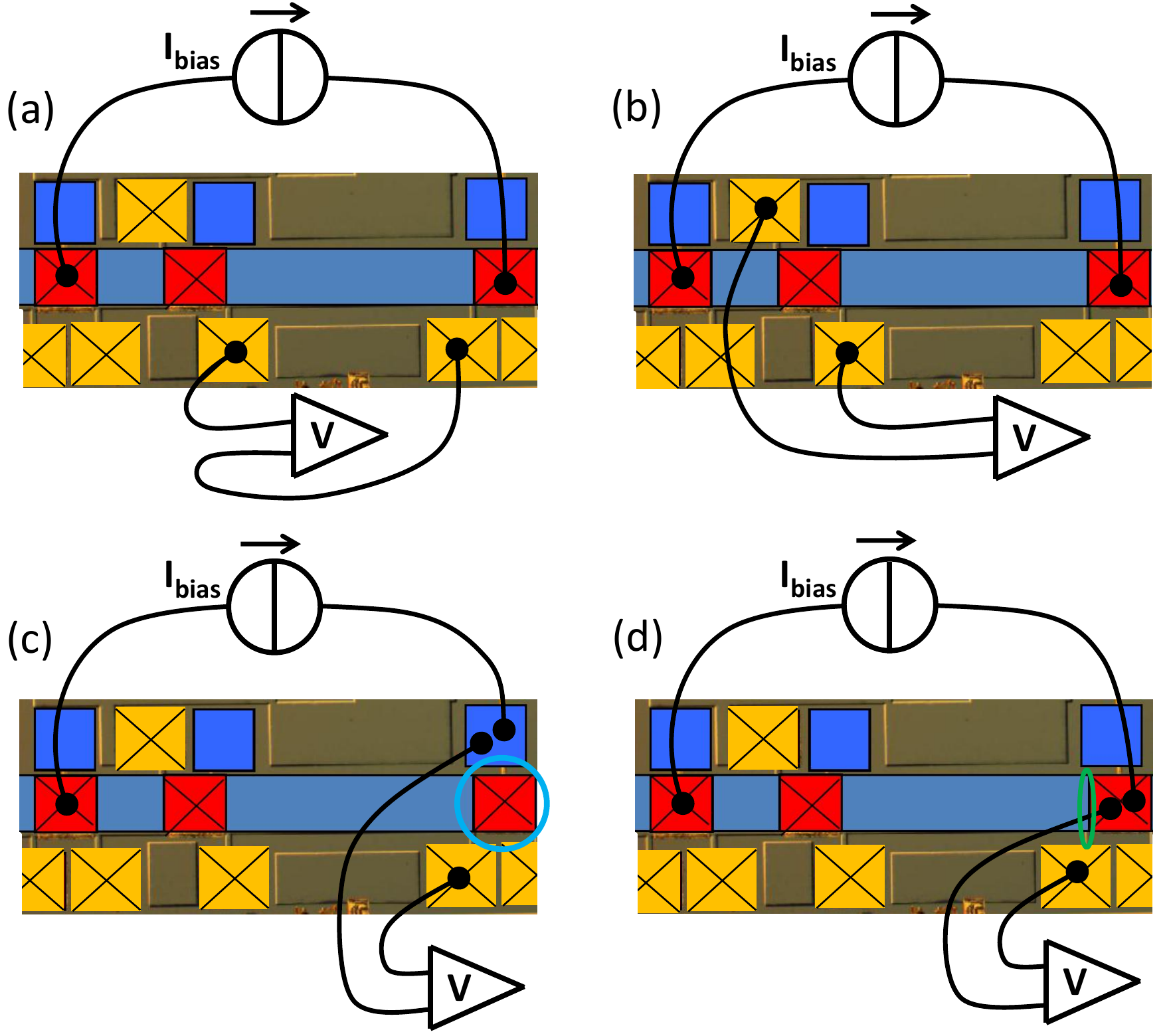}
\caption{Various measurement schemes illustrated with colored
contacts as in Fig.~\protect\ref{8Fig:DeviceScheme}. \textbf{(a)}
The scheme to measure the pristine 2DEG resistance. \textbf{(b)} The
scheme to measure the 2DEG resistance under an ohmic contact.
\textbf{(c)} The 3-point measurement scheme for measuring the
resistance of an ohmic contact without a bonding wire directly on
top of the measured contact (circled in this case). The bonding wire
is here on a metal side pad that is connected to the surface
metallization of the measured ohmic contact via a narrow metal
inter-connect. \textbf{(d)} The 3-point measurement scheme for
measuring the resistance of an ohmic contact with bonding directly
on top of the measured contact.}

\label{8Fig:MeasurementSchemes}
\end{figure}

Fig.~\ref{8Fig:TLM} and Fig.~\ref{8Fig:MeasurementSchemes} show the
various measurement schemes that we applied in this study. We first
explain the Transmission Line Method (TLM) before explaining the
other measurement schemes. The TLM method \cite{8Berger1972TLM,8Reeves1982TLM} is a very accurate method for
measuring the values of pristine 2DEG square resistance and
ohmic-contact resistance, and is widely used in research on ohmic
contacts. Fig.~\ref{8Fig:TLM} shows how the TLM method works. The
contacts are made on a 2DEG strip with an increasing distance
between pairs of adjacent contacts (Fig.~\ref{8Fig:TLM}a). The width
of the contact and channel is labeled as $W$. For resistance
measurements a four point current-biased scheme is used
(Fig.~\ref{8Fig:TLM}a). This measurement is carried out for all the
consecutive contact pairs. Fig.~\ref{8Fig:TLM}b shows a schematic of
a side view on one of the contacts and the resistance contributions
that play a role. The resistance contributions are $R_{p}$ (probe
resistance), $R_{pc}$ (probe-to-contact resistance), $R_{c}$ (actual
contact resistance between metal pad on the surface and 2DEG) and $R_{ch}$ (2DEG
resistance of the channel between the contacts).
Fig.~\ref{8Fig:TLM}c shows the corresponding circuit diagram for the
complete four-probe scheme. Since the probe ($R_{p}$) and the
probe-to-contact resistances ($R_{pc}$) are negligible as compared
to the input resistance of the voltmeter they can be neglected. The
total resistance measured between pairs of consecutive contacts is
then
\begin{equation}
R_{total} = 2R_{c}+R_{ch} = V/I, \label{8equ:RbyTLM}
\end{equation}

The plot in Fig.~\ref{8Fig:TLM}d illustrates how to extract the
contact and 2DEG square resistance values from the TLM data. The
$R_{total}$ values are plotted as a function of the channel length.
The resistance contribution $R_{ch}$ increases linearly with
increasing channel length and $R_{total}$ shows a linear dependence
with an offset from zero that is equal to $2R_{c}$. A linear fit to
the data points can thus be used to obtain $R_{c}$. In addition, the
slope of the $R_{total}$ provides an accurate measure for the square
resistance $R_{\Box}$ of pristine 2DEG (2DEG between contacts).
Using that $R_{ch}=R_{\Box}L_i/W$ this can be expressed as
$R_{\Box}=R_{c}W/L_{T}$, where the transfer length $L_{T}$ is
defined using the intercept at zero resistance for the linear trend
(see Fig.~\ref{8Fig:TLM}d). Our experiment indeed only gave results
with a linear dependence of $R_{total}$ on $L_i$ for $L_i \geqq 260
{\rm \mu m}$ (no significant deviations). While data for $L_i = 40~
{\rm \mu m}$ is candidate for showing deviations from the linear
trend (which can give insight in the resistance distribution inside
a contact), this data was discarded for the TLM analysis since it
showed large fluctuations (because it is more sensitive to the exact
alignment of the contact edges).

We used an extended TLM scheme with first measurements that used
bonding on the metal side pads (not shown in Fig.~\ref{8Fig:TLM})
and subsequently measurements that used bonding directly on top of
the ohmic contacts (as in Fig.~\ref{8Fig:TLM}) to investigate the
influence of pressing a bonding wire on an ohmic contact. Column 6
and 7 in Table~\ref{8tab:TableRes} show the contact resistance values
measured by the TLM method with bonding wires on metal side pads and
ohmic contacts respectively. We observe here $R_{c}$ values that are significantly
lower for the case with bonding directly on the ohmic contacts.
However, while TLM results give accurate results for
$R_{c}$, it is not possible to use it for measuring the 2DEG
resistance under an ohmic contact. It also does not give information on
where inside a contact the contributions to contact resistance
arise, while such information is required for detailed understanding
of the annealing mechanism, and understanding the differences between the results in column 6
and 7 in Table~\ref{8tab:TableRes}.

%%%%%%%%%%%%%%%%%%%%%%%%%%%%%%%%%%%%%%%%%%%%%%%%%%%%%%%%%%%%%%%%%%%%%%%%%%%
%TLM measurement.
%%%%%%%%%%%%%%%%%%%%%%%%%%%%%%%%%%%%%%%%%%%%%%%%%%%%%%%%%%%%%%%%%%%%%%%%%%

%%%%%%%%%%%%%%%%%%%%%%%%%%%%%%%%%%%%%%%%%%%%%%%%%%%%%%%%%%%%%%%%%%%%%%%%%%
%Measurement Scheme

%%%%%%%%%%%%%%%%%%%%%%%%%%%%%%%%%%%%%%%%%%%%%%%%%%%%%%%%%%%%%%%%%%%%%%%%%%%

We now discuss other measurement schemes that we applied for
determining the various resistance contributions. For a first round
of measurements (this order was carried out in parallel with the TLM
measurements) bonding wires were pressed on the side ohmic contacts.
Two ohmic contacts on top of the 2DEG channel are used for injecting
current into the channel (Fig.~\ref{8Fig:MeasurementSchemes}a,b).
The voltage drop across a known length of pristine 2DEG channel or
2DEG under an ohmic contacts can then be measured with the voltage
probes. The measured values $R_{\Box}$ (square resistance of
pristine 2DEG, no significant deviations from the TLM values) and
$R_{\Box, belowC}$ (square resistance of 2DEG below an ohmic
contact) are shown in column~2 and 3 of the
Table~\ref{8tab:TableRes}, respectively.

For a second round of measurements, two bonding wires were pressed
onto each metal side pad that connects to the metal layer of an
ohmic contact on top of the channel
(Fig.~\ref{8Fig:MeasurementSchemes}c). The resistance contributions
from the metal side pads and metal inter-connects (measured on each
sample, typically 10~$\Omega$) were measured separately and are
subtracted. This measurement scheme directly gives values for the
total contact resistance of contacts, which are denoted as
$R_{c,3p-pad}$ and are shown in column~4 of
Table~\ref{8tab:TableRes}.

A third round of measurements was carried out as in
Fig.~\ref{8Fig:MeasurementSchemes}d. This scheme gives directly a
value for the total contact resistance of contacts with the bonding
wires pressed directly on top of the contact that is measured. The
results are denoted as $R_{c,3p-ohm}$ and are shown in column~5 of
Table~\ref{8tab:TableRes}.

All the resistance values measured with various schemes are thus
collected in Table~\ref{8tab:TableRes} for different annealing
times. The reported values are average of (in most cases) 5
contacts. The reported error margins for column~2 to 5 are the
standard deviations of these results. For column~6 and 7, the
standard error values obtained from fitting the linear trend are
shown.

%%%%%%%%%%%%%%%%%%%%%%%%%%%%%%%%%%%%%%%%%%%%%%%%%%%%%%%%%%%%%%%%%%%%%%%%%%%%
%%% maybe use the following text in the Results and discussion section.
%The 2DEG square resistance under the contact ($R_{\Box,belowC}$)
%is lower than the pristine 2DEG square resistance
%($R_{\Box,2DEG}$) for all the annealing temperature as shown in the
%column~numbers 2 and 3 in the Table~\ref{8tab:TableRes}. Column~4 and
%5 show the contact resistance values measured with bonding wires on the
%ohmic contacts and the the metallic pads respectively. Column~ 6 and
%7 list the contact resistance values measured by TLM method with
%bonding wires on metallic pads and ohmic contacts respectively.

\section{Results and discussions}

% & jumps to next column, \\ starts new line, \hline insert horizontal line.
\begin{table}[h!]
%%\caption{Resistance values} % title of Table
\centering % used for centering table
%\renewcommand{\arraystretch}{1.2}
%{\scriptsize \
%{\footnotesize\
\resizebox{\columnwidth}{!}{
%\resizebox{\textwidth}{!}{

\begin{tabular}{|c|c|c|c|c|c|c|}
 \hline
$1$          &$2$                       &$3$                        &$4$                    &$5$                    &$6$                     &$7$\\
\hline
$t_{A}({\rm sec})$ &$R_{\Box}(\Omega)$   &$R_{\Box,belowC}(\Omega)$   &$R_{c,3p-pad}(\Omega)$ &$R_{c,3p-ohm}(\Omega)$  &$R_{c,TLM-pad}(\Omega)$ &$R_{c,TLM-ohm}(\Omega)$ \\
\hline
$-$          &Fig.~\ref{8Fig:MeasurementSchemes}a &Fig.~\ref{8Fig:MeasurementSchemes}b &Fig.~\ref{8Fig:MeasurementSchemes}c &Fig.~\ref{8Fig:MeasurementSchemes}d &Fig.~\ref{8Fig:TLM} &Fig.~\ref{8Fig:TLM}\\
\hline
20   &19.91$\pm$1.07  &2.25$\pm$1.18  &8.60$\pm$0.75  &4.49$\pm$0.36    &9.18$\pm$4.5   &0.41$\pm$3.4   \\
\hline
50   &19.61$\pm$1.56  &3.00$\pm$0.30  &5.29$\pm$1.19  &4.00$\pm$0.28    &6.28$\pm$2.86  &1.24$\pm$1.13  \\
\hline
100  &19.50$\pm$0.60  &3.17$\pm$1.55  &9.04$\pm$0.62  &4.47$\pm$0.45    &9.77$\pm$0.80  &2.38$\pm$0.41  \\
\hline
350  &20.77$\pm$0.39  &3.78$\pm$0.38  &7.76$\pm$1.02  &4.44$\pm$0.37    &not measured   &2.30$\pm$0.30  \\
\hline
\end{tabular}}
\caption{The different resistance values for different annealing
times. Row 3 list the figures with the corresponding measurement
schemes. Column~2 and 3 are for the 2DEG square resistance and 2DEG
square resistance under the contacts, respectively. Column~4 and 5
are for the contact resistance measured by the 3-point method with
the bonding wires on side pads and the ohmic contacts, respectively.
Column~6 and 7 report the contact resistance as determined with the
TLM method, with the bonding wires on side pads and the ohmic
contacts, respectively.} \label{8tab:TableRes}
\end{table}

%%%%%%%%%%%%%%%%%%%%%%%%%%%%%%%%%%%%%%%%%%%%%%%%%%%%%%%%%%%%%%%%%%%

Table~\ref{8tab:TableRes} thus lists all the measured resistance
values that were introduced. We used in total 4 bonding steps on
each device to perform the subsequent measurements (in part because
of a limited number of measurement wires in the setup). The order of
the measurements was the following: (\textit{i}) Column~2 and 3;
(\textit{ii}) Column~6; (\textit{iii}) Column~4; (\textit{iv})
Column~ 5 and 7.

The 2DEG resistance under the ohmic contacts is by about a factor 6
lower than the resistance of the pristine 2DEG (columns 2 and 3).
This occurs for all annealing times. A previous study on ohmic
contacts by G.~Sai Saravanan \textit{et al.} \cite{9Saravanan2008} and
our results \cite{IqbalSST2013} show that upon annealing Germanium
diffuses from the surface towards 2DEG, and this increases
$n$-doping near the 2DEG. While this can reduce the mobility in this
region, the effective 2DEG square resistance apparently decrease due
to the higher doping level. The resistance values are almost
constant for all the annealing times that we studied (20 to 350
sec). This suggests that Ge diffuses to 2DEG region already for
short annealing times and that the Ge concentration does not change
significantly anymore for longer annealing times.

%%% come back to this point in the later text to explain role of 30um part.

%%%%%%%%%%%%%%%%%%%%%%%%%%%%%%%%%%%%%%%%%%%%%%%%%%
% & jumps to next column, \\ starts new line, \hline insert horizontal line.
\begin{table}[h!]
%%\caption{Various expresssions for Resistance values} % title of Table
\centering % used for centering table

\resizebox{\columnwidth}{!}{

\begin{tabular}{|c|c|c|c|c|c|}
 \hline
$1$              &$2$                 &$3$                        &$4$                                                    &$5$ \\
\hline
$t_{A}~({\rm sec})$    &$R_{c,TLM-ohm}~(\Omega)$     &$R_{c}^{'}~({\rm  \Omega mm})$    &$R_{c}~({\rm \Omega cm^{2}})$            &$\rho_{bulk}~({\rm \Omega m})$ \\
\hline
20               &0.41$\pm$3.4               &0.082$\pm$0.68             &1.64$\times 10^{-4}$$\pm$1.36$\times 10^{-3}$          &0.27$\pm$2.26  \\
\hline
50               &1.24$\pm$1.13              &0.248$\pm$0.226            &4.96$\times 10^{-4}$$\pm$4.52$\times 10^{-4}$          &0.83$\pm$0.75  \\
\hline
100              &2.38$\pm$0.41              &0.476$\pm$0.82             &9.52$\times 10^{-4}$$\pm$1.64$\times 10^{-4}$          &1.59$\pm$0.27  \\
\hline
350              &2.30$\pm$0.30              &0.46$\pm$0.06              &9.2$\times 10^{-4}$$\pm$1.2$\times 10^{-4}$            &1.53$\pm$0.2   \\
\hline
\end{tabular}}
\caption{The contact resistance values as measured by the TLM method
(column~$7$ in Table~\ref{8tab:TableRes}) represented in various
forms. Column~2 shows the measured value of the contact resistance.
Columns 3, 4 and 5 show the same resistance values but converted to
a value that is normalized to the contact width (column~3), a value
for the specific contact resistance per contact area (column~4), and
a bulk resistivity value for the material in the volume between the
surface metallization and the 2DEG layer.}
\label{8tab:TableResVarious}
\end{table}
%%%%%%%%%%%%%%%%%%%%%%%%%%%%%%%%%%%%%%%%%%%%%%%%%%%%%%%%%%%%%%%%%%%

Columns~4 to 7 in Table~\ref{8tab:TableRes} show the contact
resistance values as measured with different measurement schemes.
The results show significant differences, that we can partly explain
and which provide some insight in the different contributions to the
contact resistance. As a starting point of the discussion we use the
values in column~7, which is the TLM result for directly bonding on
the ohmic contacts. This is the most unambiguous number for the
contact resistance. Column~5 lists the contact resistance values
measured with the 3-point method with bonding directly on the ohmic
contact. Column~5 has values that are typically 3~$\Omega$ higher
than the values in column~7. This can be explained by the fact that
the result of column~5 contains a series resistance contribution
from a 30~${\rm \mu m}$ wide region of pristine 2DEG (from the
distance between the ohmic contact and the 2DEG voltage probe, this
2DEG part is encircled in green in
Fig.~\ref{8Fig:MeasurementSchemes}d). The expected resistance
contribution of this part is indeed $\sim 3 \Omega$ (using
$R_{\Box}$ of column~2). This effect was also used for correcting
the values of $R_{\Box,belowC}$ in column 3.

Column~6 shows resistance values from the TLM method with bonding on
the side pad (note that 2 side pads are involved) and these results
are about $2 \times 3 ~ \Omega$ higher than the values in column~7.
Similarly, the results of column~4 (3-point, bonding on pad, note
that only 1 side pad is involved) are about $1 \times 3 ~ \Omega$
higher than the values in column~5 (3-point, bonding on ohmic). Here
we must consider two possible explanations. The first is that the
act of pressing a bond wire on top of the ohmic contact results in a
lowering of the effective contact resistance by about $3 ~ \Omega$.
The second possibility is that it results from the fact that the
metal side pads are only connected to the surface metal of the
ohmic-contact at one narrow point. This can yield that on average
the spreading resistance inside the contact gives a contribution
that is about $3 ~ \Omega$ higher for the cases with bonding on the
side pads. Given that all our measurement results and the various
contributions are on the scale of only a few $\Omega$, we can not
distinguish these cases (we could rule out that it was due to
series resistance inside the metal side pad and its inter-connect).

Our results do not allow for more detailed conclusions on the
various contributions to the contact resistance or on the annealing
mechanism. The reason is that the measured resistance values were
all much lower than expected (given our earlier work \cite{IqbalSST2013}) and that the results showed, surprisingly, almost no dependence on the annealing time. In addition, the
possible effects of spreading resistance and small series-resistance
contributions are all on the scale of a few $\Omega$, and these
values are close to the total contact resistance values and their
statistical variation. This rules out that further analysis of our
present results can give sufficient accuracy for answering the
questions that we aimed to study.

At the same time, it is an interesting result that we find very low
contact-resistance values, and that the values do almost not change
when changing the annealing time by a factor 18. In addition, these
contact resistance values rank among the lowest reported values \cite{9Jin1991}. Table~\ref{8tab:TableResVarious} provides different
representations of the resistance values that we obtained. These
values are useful for a comparison to values in the literature where
authors present values of contact resistance in various ways. When
comparing the literature one also needs to account for a dependence
on the depth of the 2DEG and the thickness of the buffer layer. Our
results on the wafer with the 2DEG at 180~nm depth (instead of
60~nm) show indeed slightly higher values, with for $R_{c,3p-pad}
\approx 15~ \Omega$. Also these samples showed almost no dependence
on annealing time (similar results for 30 sec and 550 sec).

We have at this stage little insight why the fabrication method that
we used gives such low contact-resistance values, while also being
very robust against a variation in annealing time. We have some
initial results that point out that the variation of the
heating profile as a function of time during annealing is important. For
the experiments on the samples with the 2DEG at 180~nm depth we
compared results of annealing for 550 and 600~sec annealing times
with 5~sec RTA ramp time, to results for 550~sec annealing time that
started after a 120~sec RTA ramp time (similar to the glass-tube
oven ). The samples of the latter batch had contact resistance
values that were twice as high. A second important difference with
our earlier work \cite{IqbalSST2013} is that the glass-tube oven heats the
sample in a gas flow, while the RTA heats the sample by radiation.
This can influence the exact way the surface metallization gets
heated, and thereby have an influence on the annealing mechanism.
Finally, there is possibly a role for having a suitable very clean
${\rm N_{2}}$ flow during annealing, and a very clean sample surface
before fabrication is started (samples that appeared dirty upon
inspection did not yield results with low contact-resistance
values).

We do not speculate which possible microscopic model of the ohmic contact forming could be the most appropriate one. Our present work shows that it is certain that the heterostructure right underneath the metallization is completely degenerated, either homogeneously \cite{IqbalSST2013} or in contact spikes \cite{9Taylor1994,9Taylor1998}. In the first case, the contact conductivity should be proportional the the circumference length which we cannot check due to the fixed dimensions in this study. In the latter case of contact spikes it depends on their spacing: If they are closely packed, the circumference should again determine the conductivity. If they are more apart, the area could be the leading term.
The transmission-electron-microscope studies of our earlier work \cite{IqbalSST2013} did rule out a role for spike formation. However, given the very different behavior of the annealing step we cannot conclude that this also holds for the present study.

% Main observations at the end of last paragraph

\section{Conclusions}
We developed a dedicated device to study and unravel the various
contributions to the resistance values of an ohmic contact. We could
show that the 2DEG resistance under an ohmic contact gets lower upon
annealing, and that pressing a bonding wire onto an ohmic contact
either has little influence or only lowers it by a few $\Omega$.
However, we could not fully exploit the measurement possibilities of
our device design because we obtained very low contact-resistance
values that showed no clear dependence on annealing times when these were
varied from 20 to 350 sec. Preliminary measurements show that these
results can be obtained with rapid heating (5 sec ramp time) during
annealing, and that slower ramp times cause higher contact
resistance values. A comparison with our earlier work \cite{IqbalSST2013}
also suggests that heating by radiation gives lower contact
resistance values than heating samples in a hot gas flow.

\begin{acknowledgements}
We thank B.~Wolfs, J.~Holstein and M.~de~Roosz for technical assistance. A.D.W. acknowledges financial support from the Research school Ruhr-Universit\"{a}t Bochum and the German programs BMBF Q.com-H 16KIS0109, Mercur Pr-2013-0001, and the DFH/UFA CDFA-05-06. M.J.I. acknowledges a scholarship from the Higher Education Commission of Pakistan.
\end{acknowledgements}

%% connection of current work to the area vs circ study of our earlier publication.

%%%%%%%%%%%%%%%%%%%%%%%%%%%%%%%%%%%%%%%%%%%%%%%%%%%%%%%%%%%%%%%%%%%%%%%%%
%%% REFERENCES %%%%%%%%%%%%%%%%%%%%%%%%%%%%%%%%%%%%%%%%%%%%%%%%%%%%%%%%%%
%%%%%%%%%%%%%%%%%%%%%%%%%%%%%%%%%%%%%%%%%%%%%%%%%%%%%%%%%%%%%%%%%%%%%%%%%
%\begin{references}

\clearpage

\end{document}